\documentclass[twoside,twocolumn,9pt]{article}
\usepackage{extsizes}
\usepackage[super,sort&compress,comma]{natbib} 
\usepackage[version=3]{mhchem}
\usepackage[left=1.5cm, right=1.5cm, top=1.785cm, bottom=2.0cm]{geometry}
\usepackage{balance}
\usepackage{mathptmx}
\usepackage{sectsty}
\usepackage{graphicx} 
\usepackage{lastpage}
\usepackage[format=plain,justification=justified,singlelinecheck=false,font={stretch=1.125,small,sf},labelfont=bf,labelsep=space]{caption}
\usepackage{float}
\usepackage{fancyhdr}
\usepackage{fnpos}
\usepackage[english]{babel}
\addto{\captionsenglish}{%
  
}
\usepackage{array}
\usepackage{droidsans}
\usepackage{charter}
\usepackage[T1]{fontenc}
\usepackage[usenames,dvipsnames]{xcolor}
\usepackage{setspace}
\usepackage[compact]{titlesec}
\usepackage{hyperref}

\usepackage{epstopdf}

\definecolor{cream}{RGB}{222,217,201}

\begin{document}

\pagestyle{fancy}
\thispagestyle{plain}
\fancypagestyle{plain}{
\renewcommand{\headrulewidth}{0pt}
}

\makeFNbottom
\makeatletter
\renewcommand\LARGE{\@setfontsize\LARGE{15pt}{17}}
\renewcommand\Large{\@setfontsize\Large{12pt}{14}}
\renewcommand\large{\@setfontsize\large{10pt}{12}}
\renewcommand\footnotesize{\@setfontsize\footnotesize{7pt}{10}}
\makeatother

\renewcommand{\thefootnote}{\fnsymbol{footnote}}
\renewcommand\footnoterule{\vspace*{1pt}%
\color{cream}\hrule width 3.5in height 0.4pt \color{black}\vspace*{5pt}} 
\setcounter{secnumdepth}{5}

\makeatletter 
\renewcommand\@biblabel[1]{#1}            
\renewcommand\@makefntext[1]%
{\noindent\makebox[0pt][r]{\@thefnmark\,}#1}
\makeatother 
\renewcommand{\figurename}{\small{Fig.}~}
\sectionfont{\sffamily\Large}
\subsectionfont{\normalsize}
\subsubsectionfont{\bf}
\setstretch{1.125} 
\setlength{\skip\footins}{0.8cm}
\setlength{\footnotesep}{0.25cm}
\setlength{\jot}{10pt}
\titlespacing*{\section}{0pt}{4pt}{4pt}
\titlespacing*{\subsection}{0pt}{15pt}{1pt}

\fancyfoot{}
\fancyfoot[LO,RE]{\vspace{-7.1pt}\includegraphics[height=9pt]{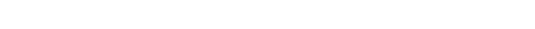}}
\fancyfoot[CO]{\vspace{-7.1pt}\hspace{11.9cm}\includegraphics{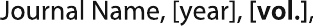}}
\fancyfoot[CE]{\vspace{-7.2pt}\hspace{-13.2cm}\includegraphics{RF}}
\fancyfoot[RO]{\footnotesize{\sffamily{1--\pageref{LastPage} ~\textbar  \hspace{2pt}\thepage}}}
\fancyfoot[LE]{\footnotesize{\sffamily{\thepage~\textbar\hspace{4.65cm} 1--\pageref{LastPage}}}}
\fancyhead{}
\renewcommand{\headrulewidth}{0pt} 
\renewcommand{\footrulewidth}{0pt}
\setlength{\arrayrulewidth}{1pt}
\setlength{\columnsep}{6.5mm}
\setlength\bibsep{1pt}

\makeatletter 
\newlength{\figrulesep} 
\setlength{\figrulesep}{0.5\textfloatsep} 

\newcommand{\topfigrule}{\vspace*{-1pt}%
\noindent{\color{cream}\rule[-\figrulesep]{\columnwidth}{1.5pt}} }

\newcommand{\botfigrule}{\vspace*{-2pt}%
\noindent{\color{cream}\rule[\figrulesep]{\columnwidth}{1.5pt}} }

\newcommand{\dblfigrule}{\vspace*{-1pt}%
\noindent{\color{cream}\rule[-\figrulesep]{\textwidth}{1.5pt}} }

\makeatother

\twocolumn[
  \begin{@twocolumnfalse}
{\includegraphics[height=30pt]{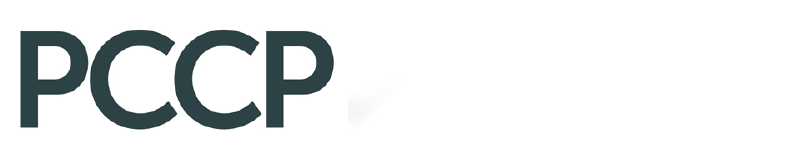}\hfill\raisebox{0pt}[0pt][0pt]{\includegraphics[height=55pt]{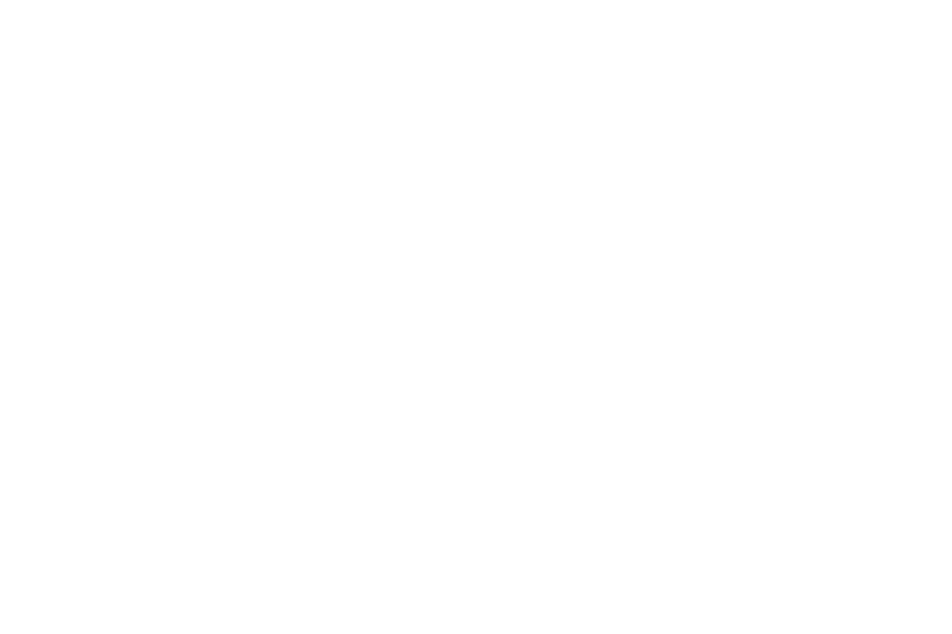}}\\[1ex]
\includegraphics[width=18.5cm]{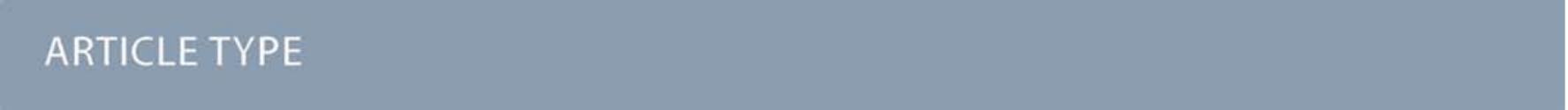}}\par
\vspace{1em}
\sffamily
\begin{tabular}{m{4.5cm} p{13.5cm} }

\includegraphics{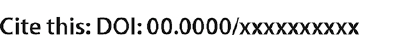} & \noindent\LARGE{\textbf{Temperature effects on the ionic conductivity in concentrated alkaline electrolyte solutions$^\dag$}} \\
 & \vspace{0.3cm} \\

 & \noindent\large{Yunqi Shao,\textit{$^{a}$} Matti Hellstr\"om,\textit{$^{b,c}$} Are Yll\"o,\textit{$^{a}$} Jonas Mindemark,\textit{$^{a}$} Kersti Hermansson,\textit{$^{a}$} J\"org Behler,\textit{$^{b}$} and Chao Zhang$^{\ast}$\textit{$^{a}$}}

\end{tabular}

 \end{@twocolumnfalse} \vspace{0.6cm}

  ]

\renewcommand*\rmdefault{bch}\normalfont\upshape
\rmfamily
\section*{}
\vspace{-1cm}


\footnotetext{\textit{$^{a}$~Department of Chemistry - \AA ngstr\"om Laboratory,
  Uppsala University, Box 538, 751 21 Uppsala, Sweden; Email: chao.zhang@kemi.uu.se}}
\footnotetext{\textit{$^{b}$~Universit\"at G\"ottingen, Institut f\"ur Physikalische Chemie, Theoretische Chemie, Tammannstr. 6, 37077 G\"ottingen, Germany }}
\footnotetext{\textit{$^{c}$~Software for Chemistry and Materials B.V., De Boelelaan 1083, 1081HV Amsterdam, The Netherlands}}
\footnotetext{\dag~Electronic Supplementary Information (ESI) available: [details of any supplementary information available should be included here]. See DOI: 00.0000/00000000.}


\rmfamily 



\textbf{Alkaline electrolyte solutions are important components in rechargeable batteries and alkaline fuel cells. As the ionic conductivity is thought to be a limiting factor in the performance of these devices, which are often operated at elevated temperatures, its temperature dependence is of significant interest. Here we use NaOH as a prototypical example of alkaline electrolytes, and for this system we have carried out reactive molecular dynamics simulations with an
  experimentally verified high-dimensional neural network
  potential derived from density-functional theory calculations.
  It is found that in concentrated NaOH solutions elevated temperatures enhance both the contributions from proton transfer to the ionic conductivity and deviations from the Nernst-Einstein relation. These findings are expected to be of practical relevance for electrochemical devices based on alkaline electrolyte solutions.}\\


Because of their excellent ionic conductivity and high room-temperature solubility, alkaline electrolyte solutions are widely used in electrochemical devices such as rechargeable batteries and alkaline fuel
cells ~\cite{Merle:2011hi, RMainar:2016ee}.  
The electrochemically active ion in alkaline electrolytes is the hydroxide ion~\cite{doi:10.1002/bbpc.19280340922}. \ce{OH-} has an anomalously high
mobility in aqueous solution, as it can diffuse via Grotthuss mechanism which is composed of a series of proton transfer events\cite{Agmon:1995ub}.
Major progress in the understanding of \ce{OH-} solvation and mobility at low concentration was made by molecular dynamics (MD)
simulations based on density functional theory~\cite{Tuckerman_1994, tuckerman_naturetransportmechanism_2002,
  marx10, hassanali_recombinationhydroniumhydroxide_2011,
  chen_hydroxidediffusesslower_2018} and reactive force fields
~\cite{day_mechanismhydratedproton_2000,
  Roberts08092009, biswas_rolepresolvationanharmonicity_2016}, which highlighted 
the importance of ``presolvation'', i.e., a thermally induced hydrogen-bond fluctuation, in the diffusion of
hydroxide ions\cite{day_mechanismhydratedproton_2000,
  tuckerman_structuredynamicsoh_2006, Roberts08092009, marx10, 
  biswas_rolepresolvationanharmonicity_2016}. 
  
Although
ionic conductivity at low concentrations is well-described by the Nernst-Einstein equation,
which links the conductivity $\sigma$ to the self-diffusion coefficients $D_+$ and $D_-$ of cations and anions, respectively, this simple picture is no longer valid at concentrations typically used
in electrochemical devices. At higher concentrations, several types of non-ideal phenomena like
ion-pairing~\cite{marcus_ionpairing_2006}, in which cations and anions associate
and form metastable neutral pairs, and, more generally, cross-correlations  of the movements of different ions (of equal or opposite charge) can notably alter the ionic conductivity. 
Moreover, the working temperature of alkaline batteries and fuel cells can be much higher than room-temperature (293K)~\cite{GUAITOLINI2018123}. Therefore, it is desirable to understand the temperature effect on proton transfer and ion-pairing in alkaline electrolyte solutions and their implications concerning the ionic conductivity at high concentrations and elevated temperatures. 

Simulations of electric properties, such as
ionic conductivity, necessitate long timescales and, except
for cases at extreme conditions~\cite{rozsa_initiospectroscopyionic_2018},  are normally beyond reach of density functional theory-based MD. One way to
tackle this time-scale challenge is to explore finite-field methods to speed
up the convergence of the polarization $\mathbf{P}$, which has been successfully
applied to compute the dielectric constant of polar liquids and the
capacitance of electrified solid-electrolyte interfaces 
\cite{Zhang:2019ce}.
The other approach to solve this problem is to make use of reactive force fields to
access longer time-scales~\cite{biswas_rolepresolvationanharmonicity_2016,Zhang:2015gz}.
One promising approach in this direction is to devise high-dimensional neural network potentials (NNPs) with density-functional theory (DFT) quality as proposed by Behler and Parrinello
\cite{Behler:2007fe}.
Here, we use this approach, and by means of molecular dynamics (MD) simulations using a NNP 
 for the prototypical case of aqueous NaOH solutions\cite{hellstrom_concentrationdependentprotontransfer_2016} and show 
how different factors in together lead to the surprising behavior of the ionic conductivity in concentrated NaOH aqueous solutions at elevated temperatures.

The details of the construction and validation of our NNP for NaOH solutions using DFT calculations at the dispersion-corrected GGA level have been discussed in Ref.~\citenum{hellstrom_concentrationdependentprotontransfer_2016}. The MD simulations were performed using LAMMPS \cite{plimpton_fastparallelalgorithms_1995} together with an extension for high-dimensional NNPs \cite{andreas__}.
The cubic simulation box contained between 272 and 496 water molecules, and between 8 and 120 NaOH formula units, depending on concentration (see Table S1 in the ESI\dag). 
The length of cubic simulation box has been fixed using the experimental densities of NaOH solutions at the given composition and temperature~\cite{green_perrychemicalengineers_1985} (see Table S1 in the ESI\dag).
Production runs with a timestep of 0.5 fs in the $NVT$ ensemble lasted for 15 ns at each combination of composition and temperature after the equilibration. The Bussi-Donadio-Parrinello thermostat~\cite{bussi_canonicalsamplingvelocity_2007} which has shown an excellent control of kinetic energy and little effect on the dynamical properties was employed.
The trajectory frames were saved every 0.01 ps for later analysis.
Each trajectory was split into 5 uncorrelated segments with length of 3 ns each. The standard deviations of observables from the different segments were used as an error estimate. Note that nuclear quantum effects were not included in the MD simulations.
 
To accompany the simulations, we have also performed conductivity measurements of NaOH solutions of concentrations up to 25 molality (m) using an "InLab" conductivity meter (Mettler Toledo). The conductivity meter probe used is a 4 pole InLab 738-ISM by (Mettler Toledo) which has a sensitivity range from 0.01--1000 mS cm$^{-1}$ and gives accurate measurements up to 373K.  The mean and the standard deviation of five independent measurements after calibration were reported for each given NaOH solution at both 293K and 323K. 

When comparing simulation and experimental results, it is important to realize that the ionic conductivity can be computed using different formulas which have different applicabilities. As mentioned at the beginning, the Nernst-Einstein equation for the ionic conductivity of a 1:1 symmetric electrolyte is valid only at low concentration and can be written as
\begin{equation}
  \label{eq:NE}
  \sigma_\text{N-E}=q^2\rho\beta(D_{+}+D_{-}) ,
\end{equation}
where $\beta$ is the inverse temperature, $q$ is the
formal charge of each ion and $\rho $ is the number density of the
formula unit of 1:1 electrolyte.

$D_{+}$ can be obtained by integrating the velocity auto-correlation function as
\begin{equation}
  \label{eq:NE_comput_vcf}
  D_+ = \frac{1}{3}\int_0^\infty \text{d}t \langle \mathbf{v}_{i,+}(0)\mathbf{v}_{i,+}(t)\rangle 
\end{equation}
where $t$ is time and $\mathbf{v}_{i,+}$ is the velocity vector of the $i$th cation, and the  average is taken over all cations and time origins. Alternatively, the Einstein
relation 
\begin{equation}
  \label{eq:NE_comput_msd}
 D_+ =   \lim_{t\to\infty} \frac{1}{6t} 
     \langle[\mathbf{r}_{i,+}(t)-\mathbf{r}_{i,+}(0)]^2\rangle 
\end{equation}
can be used, where $\mathbf{r}_{i,+}$ is the position of the $i$th cation.

$D_-$ can be computed analogously according to Eqs \ref{eq:NE_comput_vcf} and \ref{eq:NE_comput_msd}. We defined the positions of OH$^-$ ions by the position of O atoms bonded only to a single H atom. All bonds in the system were defined by assigning each hydrogen atom to its nearest oxygen atom, which gave either water molecules or hydroxide ions. Upon proton transfer reactions, the trajectories of OH$^-$ were traced traced in a fashion similar to how was done in Ref.~\citenum{hellstrom_nuclearquantumeffects_2018}, i.e. based on ``the Hungarian algorithm'' as introduced by K\"{o}nig in 1916 and Egerv\'{a}ry in 1931 and elaborated by Kuhn in 1974~\cite{Kuhn1955}. 

The Nernst-Einstein equation becomes approximate at high concentrations, for which
ion-pairing and cross-correlated ion motions play an important role~\cite{
  hansen_statisticalmechanicsdense_1975,
  zhong_selfdiffusiondistinctdiffusion_1988,
  chowdhuri_moleculardynamicssimulations_2001}. A more general equation for the conductivity, which is valid at both low and high concentration, is the Green-Kubo formula
\begin{equation}
  \label{eq:GK_comput_vcf}
  \sigma_\text{G-K} =\frac{\beta}{3}\int d\mathbf{r}\int_0^\infty
                        dt\langle\mathbf{J}(0)\mathbf{ J}(t)\rangle 
\end{equation}
where $\mathbf{J}$ is the current density. Alternatively, the equivalent Einstein-type relation can also be used~\cite{Allen:1989wd}
\begin{eqnarray}
  \label{eq:GK_comput_msd}
\sigma_\text{G-K} =
      \lim_{t\to\infty}\frac{\beta\Omega}{6t}\langle[\mathbf{P}(t)-\mathbf{P}(0)]^2\rangle 
\end{eqnarray}
 where $\mathbf{P}$ is the itinerant
polarization in ionic solution~\cite{Caillol:1994ho} and $\Omega$ is the volume of the simulation box. 

Unlike $\sigma_{\text{N-E}}$, $\sigma_{\text{G-K}}$
includes ion-pairing and cross-correlated ion motions from the so-called distinct diffusion coefficients of cations $D_{+}^{\text{d}}$,
anions $D_{-}^{\text{d}}$ and cation-anion pairs
$D_{+-}^{\text{d}}$,
which can be computed from MD simulations~\cite{zhong_selfdiffusiondistinctdiffusion_1988,kashyap_howchargetransport_2011}. The name ``distinct'' means it is cross-correlation between two different ions, even within the same species. 


This leads to a decomposition of the Green-Kubo conductivity as
\begin{eqnarray}
  \sigma_\text{G-K} &=& q^2\rho\beta(D_{+}+D_{-} + D_{+}^{\text{d}}/2 +
                        D_{-}^{\text{d}}/2  - D_{+-}^ {\text{d}})  \\
  \label{eq:GK_distinct}
                    &=& \sigma_\text{N-E} + \sigma_{+}^{\text{d}} +
                        \sigma_{-}^{\text{d}} + \sigma_{+-}^{\text{d}} 
\end{eqnarray} 
where $\sigma_{+}^{\text{d}} $, $\sigma_{-}^{\text{d}}$ and
$\sigma_{+-}^{\text{d}} $ are contributions to the ionic
conductivity from the corresponding distinct diffusion coefficients.

\begin{figure}
  \includegraphics[width=1.0\columnwidth]{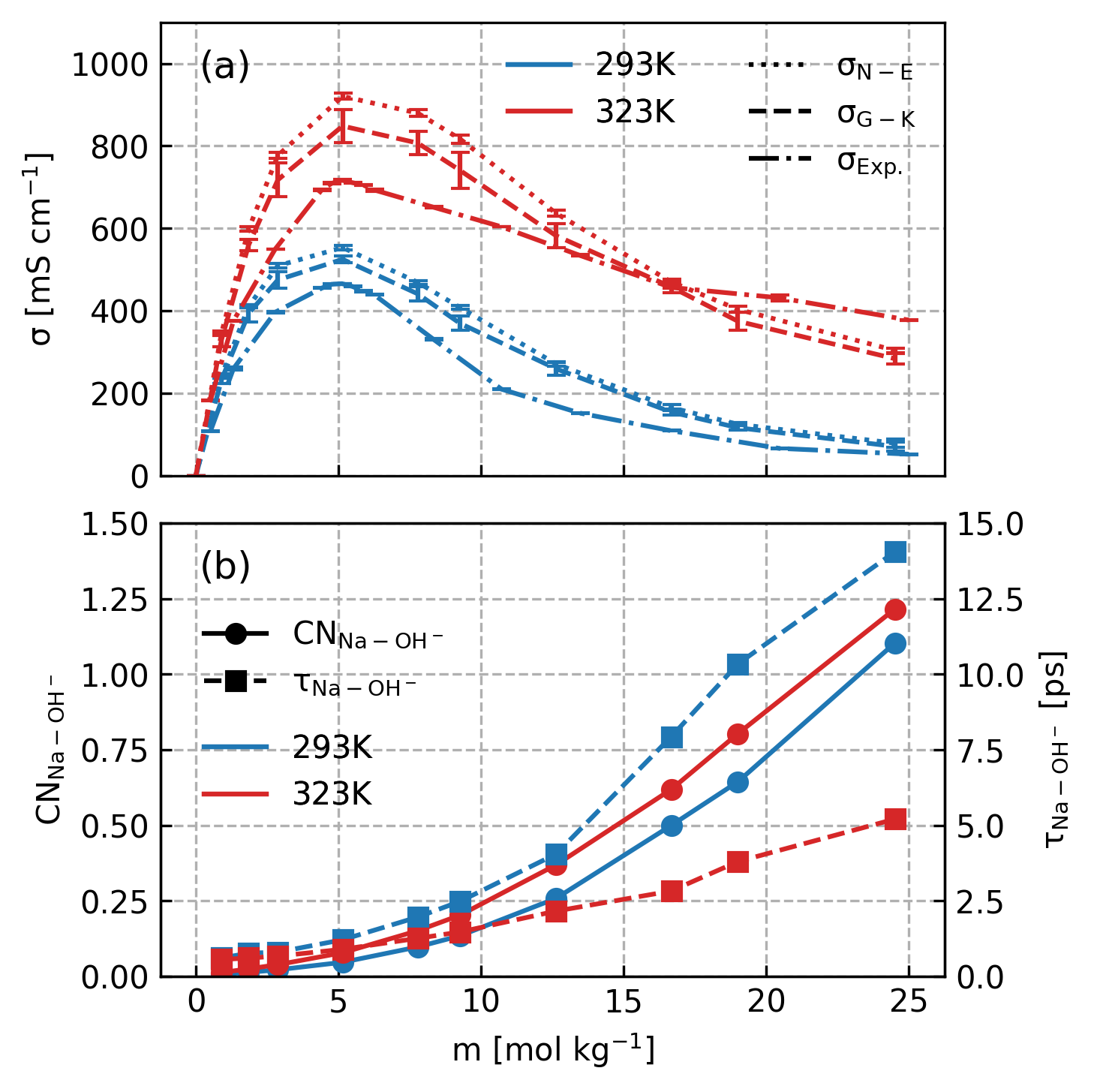}
  \caption{
  (a) Comparison of concentration-dependent ionic conductivities calculated using the Nernst-Einstein formula (Eq.~\ref{eq:NE} and
  Eq.~\ref{eq:NE_comput_msd}) and the Green-Kubo formula (Eq.~\ref{eq:GK_comput_msd}) from MD simulations
  and those measured in experiments at 293K and 323K;
  (b) Calculated Na$^+-$OH$^-$ coordination numbers and residence time.}
  \label{fgr:cn_cond_compare}
\end{figure}

\begin{figure*}[h]
  \includegraphics[width=2.0\columnwidth]{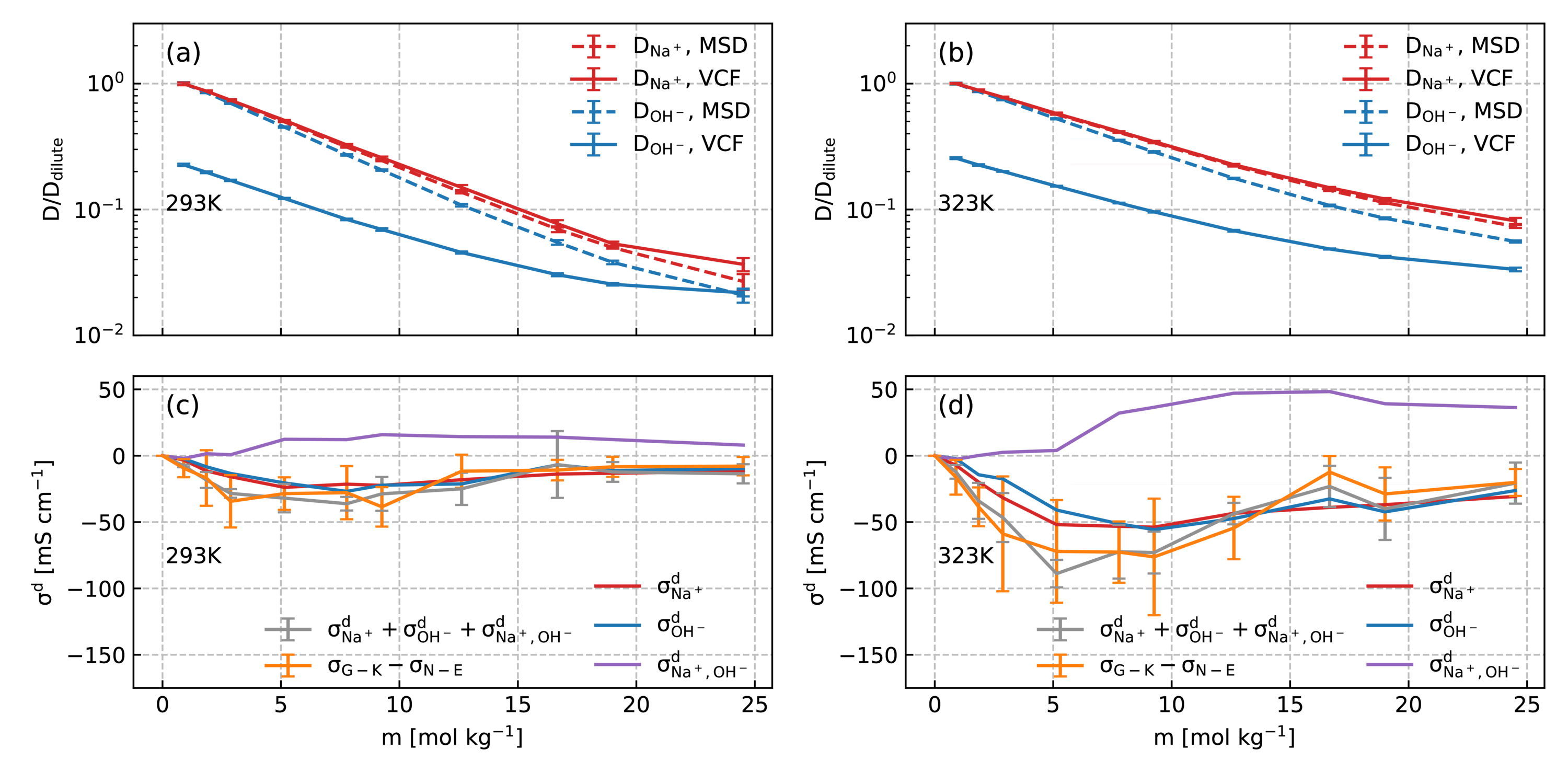}
  \caption{(a) and (b) Scaled concentration-dependent self-diffusion coefficients of Na$^+$ and OH$^-$ at 293K and 323K obtained
    from VCF (Eq.~\ref{eq:NE_comput_vcf}) and
    from MSD (Eq.~\ref{eq:NE_comput_msd}),
    where the self-diffusion coefficients were scaled by the corresponding value
    in the dilute solution; (c) and (d) Contributions from distinct diffusion coefficients to the ionic conductivity
    at 293K and 323K as shown in Eq.~\ref{eq:GK_distinct} and calculated from VCFs. Detailed definition of distinct diffusion coefficients can be found in Ref.\citenum{zhong_selfdiffusiondistinctdiffusion_1988,kashyap_howchargetransport_2011}.}
  \label{fgr:vcf_cond_compare}
\end{figure*}
After elaborating on the difference between the Nernst-Einstein conductivity and the Green-Kubo conductivity, we are now ready to compare the ionic conductivity calculated from MD simulations to those measured in experiments. The results are shown in Fig.~\ref{fgr:cn_cond_compare}. In Fig.~\ref{fgr:cn_cond_compare}a, the calculated ionic conductivities from MD simulations agree well with conductivity measurements, especially at 293K. Considering that the NNP~\cite{hellstrom_concentrationdependentprotontransfer_2016} was generated using only DFT calculations as the reference, this agreement is quite encouraging. Note that we have neglected the finite-size correction \cite{Yeh2004} to the Nernst-Einstein conductivity showed in Fig.~\ref{fgr:cn_cond_compare}a,  because of the relatively large simulation box that we used and the high viscosity of concentrated NaOH solutions (See Fig. S1 in the ESI\dag). 

When inspecting the simulation results at both 293K and 323K (Fig.~\ref{fgr:cn_cond_compare}), one can see that the Nernst-Einstein conductivity is always larger than the Green-Kubo conductivity, as expected. However, there are several interesting observations specific to NaOH solutions. First, the absolute difference between the Nernst-Einstein conductivity and the Green-Kubo conductivity becomes smaller at higher concentrations which is counter-intuitive. 
Second, the difference is in general larger at higher temperature (323K) than at lower temperature (293K). Third, near the solubility limit (25 m) at  room-temperature, the ionic conductivity at 323K is still substantial while at 293K it becomes quite small. 

One way to rationalize these observations is to consider ion-pairing. For this reason, we calculated the coordination number of OH$^-$ around Na$^+$ ions as well the residence time of Na$^+-$OH$^-$ pairs. The coordination number was calculated by integrating the radial distribution function to its first minimum and the residence time was calculated by following the stable states picture formalism~\cite{laage2008}.
As is described in Ref~\citenum{Hellstrom:2017gf}, a time correlation function $C(t)$ was calculated to give the probability that a "stable" hydroxide ion at time $t$ does not escape the first coordination shell of Na$^+$ through either ligand exchange or proton transfer within the interval $t_0$ and $t_0 + t$. 
$C(t)$ was then fitted to a biexponentially decaying function to extract the residence time. As shown in Fig.~\ref{fgr:cn_cond_compare}b, the number of OH$^-$ coordinating Na$^+$ at both 293K and 323K can exceed one near the room-temperature solubility limit. However, the residence time at 293K increases much more rapidly with the concentration than that at 323K. Base on these observations, we speculate that the effect of ion-pairing on the ionic conductivity would be stronger at 293K.

In order to dissect contributions from proton transfer reactions and cross-correlated ion motions, we exploited the fact that proton transfer contributes to the
mean squared displacement (MSD, Eq.~\ref{eq:NE_comput_msd}) but not to the corresponding
velocity correlation function (VCF, Eq.~\ref{eq:NE_comput_vcf}).
As shown in Fig.~\ref{fgr:vcf_cond_compare}a and Fig.~\ref{fgr:vcf_cond_compare}b, the scaled self-diffusion coefficients of Na$^+$ are the same regardless whether they are computed from MSD or from VCF. In contrast, the scaled self-diffusion coefficients of OH$^-$ computed from MSD and VCF are different and their difference quantifies the contribution from proton transfer reactions. Comparing to the case at 323K, one can clearly see that the proton transfer contribution to the self-diffusion coefficient of OH$^-$ becomes negligible at 293K in concentrated NaOH solutions. This is the main reason why the ionic conductivity near the room-temperature solubility limit at 323K is still significant while  at 293K it becomes severely diminished. 

We then used the same technique to evaluate the contributions from 
the distinct diffusion coefficients to the ionic conductivity from VCFs, as shown in Fig.~\ref{fgr:vcf_cond_compare}c and Fig.~\ref{fgr:vcf_cond_compare}d. It is found that $\sigma_{\textrm{Na}^+, \textrm{OH}^-}^{\text{d}} $ at both 293K and 323K are positive instead of negative as a simple picture of ion-pairing would suggest. Further, $\sigma_{\textrm{Na}^+, \textrm{OH}^-}^{\text{d}} $ is much more positive at 323 K than that at 293 K. If one considers that the negativeness of $\sigma_{\textrm{Na}^+, \textrm{OH}^-}^{\text{d}}$ as the onset of ion-pairing contributions to the ionic conductivity, then the effect of ion-pairing on the ionic conductivity are more likely to happen at 293 K than that at 323 K. 

When comparing the sum of $\sigma_{\textrm{Na}^+}^{\text{d}} $, $\sigma_{\textrm{OH}^-}^{\text{d}}$ and $\sigma_{\textrm{Na}^+, \textrm{OH}^-}^{\text{d}} $ calculated from VCFs to the absolute difference between $\sigma_\textrm{G-K}$ and $\sigma_\textrm{N-E}$ calculated from the MSD, one can see a good agreement within the statistical error. Note that $(\sigma_\textrm{G-K}-\sigma_\textrm{N-E})$ has a larger error bar than its counterpart $(\sigma_{\textrm{Na}^+}^{\text{d}}+ \sigma_{\textrm{OH}^-}^{\text{d}} + \sigma_{\textrm{Na}^+, \textrm{OH}^-}^{\text{d}})$, because the former is the difference of two large numbers (Fig.~\ref{fgr:cn_cond_compare}a). The agreement between these two quantities means that the cross-correlated ion motions are hydrodynamic in nature and not determined by proton transfer reactions in NaOH solutions. We suspect that the counter-intuitive observation that $(\sigma_\textrm{G-K}-\sigma_\textrm{N-E})$ becomes smaller at high concentration is related to the rapid increment of the viscosity in NaOH solutions (See Fig. S2 in the ESI\dag). Since the viscosity decreases at elevated temperatures, this may also explain a larger difference between the Nernst-Einstein conductivity and the Green-Kubo conductivity at 323K as seen in Fig.~\ref{fgr:cn_cond_compare}a. Nevertheless, future investigations are needed to identify the factors affecting the cross-correlated ion motions and subsequent deviations from the Nernst-Einstein relation~\cite{Shao2019}. 

\section*{Conflicts of interest}
There are no conflicts to declare.

\section*{Acknowledgements}
CZ thanks Uppsala University (UU) for a start-up grant and funding from the
  Swedish National Strategic e-Science programme eSSENCE.
  MH acknowledges funding from the European Union's Horizon 2020 research and innovation programme under grant agreement No 798129. JB acknowledges the DFG for a DFG Heisenberg professorship (Be3264/11-2, project 329898176).
  The simulations were performed on the resources provided by the Swedish National Infrastructure for Computing (SNIC) at NSC.



\begin{mcitethebibliography}{36}
\providecommand*{\natexlab}[1]{#1}
\providecommand*{\mciteSetBstSublistMode}[1]{}
\providecommand*{\mciteSetBstMaxWidthForm}[2]{}
\providecommand*{\mciteBstWouldAddEndPuncttrue}
  {\def\EndOfBibitem{\unskip.}}
\providecommand*{\mciteBstWouldAddEndPunctfalse}
  {\let\EndOfBibitem\relax}
\providecommand*{\mciteSetBstMidEndSepPunct}[3]{}
\providecommand*{\mciteSetBstSublistLabelBeginEnd}[3]{}
\providecommand*{\EndOfBibitem}{}
\mciteSetBstSublistMode{f}
\mciteSetBstMaxWidthForm{subitem}
{(\emph{\alph{mcitesubitemcount}})}
\mciteSetBstSublistLabelBeginEnd{\mcitemaxwidthsubitemform\space}
{\relax}{\relax}

\bibitem[Merle \emph{et~al.}(2011)Merle, Wessling, and Nijmeijer]{Merle:2011hi}
G.~Merle, M.~Wessling and K.~Nijmeijer, \emph{J. Membr. Sci.}, 2011,
  \textbf{377}, 1--35\relax
\mciteBstWouldAddEndPuncttrue
\mciteSetBstMidEndSepPunct{\mcitedefaultmidpunct}
{\mcitedefaultendpunct}{\mcitedefaultseppunct}\relax
\EndOfBibitem
\bibitem[R~Mainar \emph{et~al.}(2016)R~Mainar, Leonet, Bengoechea, Boyano,
  de~Meatza, Kvasha, Guerfi, and Alberto~Bl{\'a}zquez]{RMainar:2016ee}
A.~R~Mainar, O.~Leonet, M.~Bengoechea, I.~Boyano, I.~de~Meatza, A.~Kvasha,
  A.~Guerfi and J.~Alberto~Bl{\'a}zquez, \emph{Int. J. Energy Res.}, 2016,
  \textbf{40}, 1032--1049\relax
\mciteBstWouldAddEndPuncttrue
\mciteSetBstMidEndSepPunct{\mcitedefaultmidpunct}
{\mcitedefaultendpunct}{\mcitedefaultseppunct}\relax
\EndOfBibitem
\bibitem[H{\"u}ckel(1928)]{doi:10.1002/bbpc.19280340922}
E.~H{\"u}ckel, \emph{Z. Elektrochem. Angew. Physik. Chem.}, 1928, \textbf{34},
  546--562\relax
\mciteBstWouldAddEndPuncttrue
\mciteSetBstMidEndSepPunct{\mcitedefaultmidpunct}
{\mcitedefaultendpunct}{\mcitedefaultseppunct}\relax
\EndOfBibitem
\bibitem[Agmon(1995)]{Agmon:1995ub}
N.~Agmon, \emph{Chem. Phys. Lett.}, 1995, \textbf{244}, 456--462\relax
\mciteBstWouldAddEndPuncttrue
\mciteSetBstMidEndSepPunct{\mcitedefaultmidpunct}
{\mcitedefaultendpunct}{\mcitedefaultseppunct}\relax
\EndOfBibitem
\bibitem[Tuckerman \emph{et~al.}(1994)Tuckerman, Laasonen, Sprik, and
  Parrinello]{Tuckerman_1994}
M.~E. Tuckerman, K.~Laasonen, M.~Sprik and M.~Parrinello, \emph{J. Phys.:
  Condens. Mat.}, 1994, \textbf{6}, A93--A100\relax
\mciteBstWouldAddEndPuncttrue
\mciteSetBstMidEndSepPunct{\mcitedefaultmidpunct}
{\mcitedefaultendpunct}{\mcitedefaultseppunct}\relax
\EndOfBibitem
\bibitem[Tuckerman \emph{et~al.}(2002)Tuckerman, Marx, and
  Parrinello]{tuckerman_naturetransportmechanism_2002}
M.~E. Tuckerman, D.~Marx and M.~Parrinello, \emph{Nature}, 2002, \textbf{417},
  925--929\relax
\mciteBstWouldAddEndPuncttrue
\mciteSetBstMidEndSepPunct{\mcitedefaultmidpunct}
{\mcitedefaultendpunct}{\mcitedefaultseppunct}\relax
\EndOfBibitem
\bibitem[Marx \emph{et~al.}(2010)Marx, Chandra, and Tuckerman]{marx10}
D.~Marx, A.~Chandra and M.~E. Tuckerman, \emph{Chem. Rev.}, 2010, \textbf{110},
  2174--2216\relax
\mciteBstWouldAddEndPuncttrue
\mciteSetBstMidEndSepPunct{\mcitedefaultmidpunct}
{\mcitedefaultendpunct}{\mcitedefaultseppunct}\relax
\EndOfBibitem
\bibitem[Hassanali \emph{et~al.}(2011)Hassanali, Prakash, Eshet, and
  Parrinello]{hassanali_recombinationhydroniumhydroxide_2011}
A.~Hassanali, M.~K. Prakash, H.~Eshet and M.~Parrinello, \emph{Proc. Natl.
  Acad. Sci. U.S.A.}, 2011, \textbf{108}, 20410--20415\relax
\mciteBstWouldAddEndPuncttrue
\mciteSetBstMidEndSepPunct{\mcitedefaultmidpunct}
{\mcitedefaultendpunct}{\mcitedefaultseppunct}\relax
\EndOfBibitem
\bibitem[Chen \emph{et~al.}(2018)Chen, Zheng, Santra, Ko, Jr, Klein, Car, and
  Wu]{chen_hydroxidediffusesslower_2018}
M.~Chen, L.~Zheng, B.~Santra, H.-Y. Ko, R.~A.~D. Jr, M.~L. Klein, R.~Car and
  X.~Wu, \emph{Nat. Chem.}, 2018, \textbf{10}, 413--419\relax
\mciteBstWouldAddEndPuncttrue
\mciteSetBstMidEndSepPunct{\mcitedefaultmidpunct}
{\mcitedefaultendpunct}{\mcitedefaultseppunct}\relax
\EndOfBibitem
\bibitem[Day \emph{et~al.}(2000)Day, Schmitt, and
  Voth]{day_mechanismhydratedproton_2000}
T.~J.~F. Day, U.~W. Schmitt and G.~A. Voth, \emph{J. Am. Chem. Soc.}, 2000,
  \textbf{122}, 12027--12028\relax
\mciteBstWouldAddEndPuncttrue
\mciteSetBstMidEndSepPunct{\mcitedefaultmidpunct}
{\mcitedefaultendpunct}{\mcitedefaultseppunct}\relax
\EndOfBibitem
\bibitem[Roberts \emph{et~al.}(2009)Roberts, Petersen, Ramasesha, Tokmakoff,
  Ufimtsev, and Martinez]{Roberts08092009}
S.~T. Roberts, P.~B. Petersen, K.~Ramasesha, A.~Tokmakoff, I.~S. Ufimtsev and
  T.~J. Martinez, \emph{Proc. Natl. Acad. Sci. U.S.A.}, 2009, \textbf{106},
  15154--15159\relax
\mciteBstWouldAddEndPuncttrue
\mciteSetBstMidEndSepPunct{\mcitedefaultmidpunct}
{\mcitedefaultendpunct}{\mcitedefaultseppunct}\relax
\EndOfBibitem
\bibitem[Biswas \emph{et~al.}(2016)Biswas, Tse, Tokmakoff, and
  Voth]{biswas_rolepresolvationanharmonicity_2016}
R.~Biswas, Y.-L.~S. Tse, A.~Tokmakoff and G.~A. Voth, \emph{J. Phys. Chem. B},
  2016, \textbf{120}, 1793--1804\relax
\mciteBstWouldAddEndPuncttrue
\mciteSetBstMidEndSepPunct{\mcitedefaultmidpunct}
{\mcitedefaultendpunct}{\mcitedefaultseppunct}\relax
\EndOfBibitem
\bibitem[Tuckerman \emph{et~al.}(2006)Tuckerman, Chandra, and
  Marx]{tuckerman_structuredynamicsoh_2006}
M.~E. Tuckerman, A.~Chandra and D.~Marx, \emph{Acc. Chem. Res.}, 2006,
  \textbf{39}, 151--158\relax
\mciteBstWouldAddEndPuncttrue
\mciteSetBstMidEndSepPunct{\mcitedefaultmidpunct}
{\mcitedefaultendpunct}{\mcitedefaultseppunct}\relax
\EndOfBibitem
\bibitem[Marcus and Hefter(2006)]{marcus_ionpairing_2006}
Y.~Marcus and G.~Hefter, \emph{Chem. Rev.}, 2006, \textbf{106},
  4585--4621\relax
\mciteBstWouldAddEndPuncttrue
\mciteSetBstMidEndSepPunct{\mcitedefaultmidpunct}
{\mcitedefaultendpunct}{\mcitedefaultseppunct}\relax
\EndOfBibitem
\bibitem[Guaitolini and Fardin(2018)]{GUAITOLINI2018123}
S.~V. Guaitolini and J.~F. Fardin, \emph{Advances in Renewable Energies and
  Power Technologies}, Elsevier, 2018, pp. 123 -- 150\relax
\mciteBstWouldAddEndPuncttrue
\mciteSetBstMidEndSepPunct{\mcitedefaultmidpunct}
{\mcitedefaultendpunct}{\mcitedefaultseppunct}\relax
\EndOfBibitem
\bibitem[Rozsa \emph{et~al.}(2018)Rozsa, Pan, Giberti, and
  Galli]{rozsa_initiospectroscopyionic_2018}
V.~Rozsa, D.~Pan, F.~Giberti and G.~Galli, \emph{Proc. Natl. Acad. Sci.
  U.S.A.}, 2018, \textbf{115}, 6952--6957\relax
\mciteBstWouldAddEndPuncttrue
\mciteSetBstMidEndSepPunct{\mcitedefaultmidpunct}
{\mcitedefaultendpunct}{\mcitedefaultseppunct}\relax
\EndOfBibitem
\bibitem[Zhang \emph{et~al.}(2019)Zhang, Hutter, and Sprik]{Zhang:2019ce}
C.~Zhang, J.~Hutter and M.~Sprik, \emph{J. Phys. Chem. Lett.}, 2019,
  \textbf{10}, 3871--3876\relax
\mciteBstWouldAddEndPuncttrue
\mciteSetBstMidEndSepPunct{\mcitedefaultmidpunct}
{\mcitedefaultendpunct}{\mcitedefaultseppunct}\relax
\EndOfBibitem
\bibitem[Zhang and van Duin(2015)]{Zhang:2015gz}
W.~Zhang and A.~C.~T. van Duin, \emph{J. Phys. Chem. C}, 2015, \textbf{119},
  27727--27736\relax
\mciteBstWouldAddEndPuncttrue
\mciteSetBstMidEndSepPunct{\mcitedefaultmidpunct}
{\mcitedefaultendpunct}{\mcitedefaultseppunct}\relax
\EndOfBibitem
\bibitem[Behler and Parrinello(2007)]{Behler:2007fe}
J.~Behler and M.~Parrinello, \emph{Phys. Rev. Lett.}, 2007, \textbf{98},
  146401\relax
\mciteBstWouldAddEndPuncttrue
\mciteSetBstMidEndSepPunct{\mcitedefaultmidpunct}
{\mcitedefaultendpunct}{\mcitedefaultseppunct}\relax
\EndOfBibitem
\bibitem[Hellstr\"om and
  Behler(2016)]{hellstrom_concentrationdependentprotontransfer_2016}
M.~Hellstr\"om and J.~Behler, \emph{J. Phys. Chem. Lett.}, 2016, \textbf{7},
  3302--3306\relax
\mciteBstWouldAddEndPuncttrue
\mciteSetBstMidEndSepPunct{\mcitedefaultmidpunct}
{\mcitedefaultendpunct}{\mcitedefaultseppunct}\relax
\EndOfBibitem
\bibitem[Plimpton(1995)]{plimpton_fastparallelalgorithms_1995}
S.~Plimpton, \emph{J. Comput. Phys.}, 1995, \textbf{117}, 1--19\relax
\mciteBstWouldAddEndPuncttrue
\mciteSetBstMidEndSepPunct{\mcitedefaultmidpunct}
{\mcitedefaultendpunct}{\mcitedefaultseppunct}\relax
\EndOfBibitem
\bibitem[Singraber \emph{et~al.}(2019)Singraber, Behler, and
  Dellago]{andreas__}
A.~Singraber, J.~Behler and C.~Dellago, \emph{J. Chem. Theory Comput.}, 2019,
  \textbf{15}, 1827--1840\relax
\mciteBstWouldAddEndPuncttrue
\mciteSetBstMidEndSepPunct{\mcitedefaultmidpunct}
{\mcitedefaultendpunct}{\mcitedefaultseppunct}\relax
\EndOfBibitem
\bibitem[Perry \emph{et~al.}(1984)Perry, Green, and
  Maloney]{green_perrychemicalengineers_1985}
R.~H. Perry, D.~W. Green and J.~O. Maloney, \emph{Perry's {Chemical} engineers'
  handbook}, McGraw-Hill, New York, 6th edn, 1984\relax
\mciteBstWouldAddEndPuncttrue
\mciteSetBstMidEndSepPunct{\mcitedefaultmidpunct}
{\mcitedefaultendpunct}{\mcitedefaultseppunct}\relax
\EndOfBibitem
\bibitem[Bussi \emph{et~al.}(2007)Bussi, Donadio, and
  Parrinello]{bussi_canonicalsamplingvelocity_2007}
G.~Bussi, D.~Donadio and M.~Parrinello, \emph{J. Chem. Phys.}, 2007,
  \textbf{126}, 014101\relax
\mciteBstWouldAddEndPuncttrue
\mciteSetBstMidEndSepPunct{\mcitedefaultmidpunct}
{\mcitedefaultendpunct}{\mcitedefaultseppunct}\relax
\EndOfBibitem
\bibitem[Hellstr{\"o}m \emph{et~al.}(2018)Hellstr{\"o}m, Ceriotti, and
  Behler]{hellstrom_nuclearquantumeffects_2018}
M.~Hellstr{\"o}m, M.~Ceriotti and J.~Behler, \emph{J. Phys. Chem. B.}, 2018,
  \textbf{122}, 10158--10171\relax
\mciteBstWouldAddEndPuncttrue
\mciteSetBstMidEndSepPunct{\mcitedefaultmidpunct}
{\mcitedefaultendpunct}{\mcitedefaultseppunct}\relax
\EndOfBibitem
\bibitem[Kuhn(1955)]{Kuhn1955}
H.~W. Kuhn, \emph{Nav. Res. Logist. Q.}, 1955, \textbf{2}, 83--97\relax
\mciteBstWouldAddEndPuncttrue
\mciteSetBstMidEndSepPunct{\mcitedefaultmidpunct}
{\mcitedefaultendpunct}{\mcitedefaultseppunct}\relax
\EndOfBibitem
\bibitem[Hansen and McDonald(1975)]{hansen_statisticalmechanicsdense_1975}
J.~P. Hansen and I.~R. McDonald, \emph{Phys. Rev. A}, 1975, \textbf{11},
  2111--2123\relax
\mciteBstWouldAddEndPuncttrue
\mciteSetBstMidEndSepPunct{\mcitedefaultmidpunct}
{\mcitedefaultendpunct}{\mcitedefaultseppunct}\relax
\EndOfBibitem
\bibitem[Zhong and Friedman(1988)]{zhong_selfdiffusiondistinctdiffusion_1988}
E.~C. Zhong and H.~L. Friedman, \emph{J. Phys. Chem.}, 1988, \textbf{92},
  1685--1692\relax
\mciteBstWouldAddEndPuncttrue
\mciteSetBstMidEndSepPunct{\mcitedefaultmidpunct}
{\mcitedefaultendpunct}{\mcitedefaultseppunct}\relax
\EndOfBibitem
\bibitem[Chowdhuri and
  Chandra(2001)]{chowdhuri_moleculardynamicssimulations_2001}
S.~Chowdhuri and A.~Chandra, \emph{J. Chem. Phys.}, 2001, \textbf{115},
  3732--3741\relax
\mciteBstWouldAddEndPuncttrue
\mciteSetBstMidEndSepPunct{\mcitedefaultmidpunct}
{\mcitedefaultendpunct}{\mcitedefaultseppunct}\relax
\EndOfBibitem
\bibitem[Allen and Tildesley(2017)]{Allen:1989wd}
M.~P. Allen and D.~J. Tildesley, \emph{Computer simulation of liquids}, Oxford
  university press, 2017\relax
\mciteBstWouldAddEndPuncttrue
\mciteSetBstMidEndSepPunct{\mcitedefaultmidpunct}
{\mcitedefaultendpunct}{\mcitedefaultseppunct}\relax
\EndOfBibitem
\bibitem[Caillol(1994)]{Caillol:1994ho}
J.-M. Caillol, \emph{J. Chem. Phys.}, 1994, \textbf{101}, 6080--6090\relax
\mciteBstWouldAddEndPuncttrue
\mciteSetBstMidEndSepPunct{\mcitedefaultmidpunct}
{\mcitedefaultendpunct}{\mcitedefaultseppunct}\relax
\EndOfBibitem
\bibitem[Kashyap \emph{et~al.}(2011)Kashyap, Annapureddy, Raineri, and
  Margulis]{kashyap_howchargetransport_2011}
H.~K. Kashyap, H.~V.~R. Annapureddy, F.~O. Raineri and C.~J. Margulis, \emph{J.
  Phys. Chem. B}, 2011, \textbf{115}, 13212--13221\relax
\mciteBstWouldAddEndPuncttrue
\mciteSetBstMidEndSepPunct{\mcitedefaultmidpunct}
{\mcitedefaultendpunct}{\mcitedefaultseppunct}\relax
\EndOfBibitem
\bibitem[Yeh and Hummer(2004)]{Yeh2004}
I.-C. Yeh and G.~Hummer, \emph{J. Phys. Chem. B}, 2004, \textbf{108},
  15873--15879\relax
\mciteBstWouldAddEndPuncttrue
\mciteSetBstMidEndSepPunct{\mcitedefaultmidpunct}
{\mcitedefaultendpunct}{\mcitedefaultseppunct}\relax
\EndOfBibitem
\bibitem[Laage and Hynes(2008)]{laage2008}
D.~Laage and J.~T. Hynes, \emph{J. Phys. Chem. B}, 2008, \textbf{112},
  7697--7701\relax
\mciteBstWouldAddEndPuncttrue
\mciteSetBstMidEndSepPunct{\mcitedefaultmidpunct}
{\mcitedefaultendpunct}{\mcitedefaultseppunct}\relax
\EndOfBibitem
\bibitem[Hellstr{\"o}m and Behler(2017)]{Hellstrom:2017gf}
M.~Hellstr{\"o}m and J.~Behler, \emph{J. Phys. Chem. B}, 2017, \textbf{121},
  4184--4190\relax
\mciteBstWouldAddEndPuncttrue
\mciteSetBstMidEndSepPunct{\mcitedefaultmidpunct}
{\mcitedefaultendpunct}{\mcitedefaultseppunct}\relax
\EndOfBibitem
\bibitem[Shao \emph{et~al.}(2019)Shao, Shigenobu, Watanabe, and
  Zhang]{Shao2019}
Y.~Shao, K.~Shigenobu, M.~Watanabe and C.~Zhang, \emph{chemRxiv.8217152.v2},
  2019\relax
\mciteBstWouldAddEndPuncttrue
\mciteSetBstMidEndSepPunct{\mcitedefaultmidpunct}
{\mcitedefaultendpunct}{\mcitedefaultseppunct}\relax
\EndOfBibitem
\end{mcitethebibliography}
  \providecommand*{\mcitethebibliography}{\thebibliography}
\csname @ifundefined\endcsname{endmcitethebibliography}
{\let\endmcitethebibliography\endthebibliography}{}

\bibliographystyle{rsc} 

\end{document}


\section{Finite-size correction to the Nernst-Einstein conductivity}
As is discussed in the main text, the Nernst-Einstein conductivity is computed from the 
self-diffusion coefficients of cation $D_+$ and anion $D_-$ (Eq.~\ref{eq:NE}).

\begin{equation}
   \sigma_\text{N-E}=q^2\rho\beta(D_{+}+D_{-})
    \label{eq:NE}
\end{equation}

The self-diffusion coefficients with periodic boundary conditions are known to have system-size dependence\cite{Yeh2004}.
The dependence is mainly due to the periodicity-induced hydrodynamic self-interaction which
can be corrected using Eq.~\ref{eq:correction}, where 
$D_0$ is corrected self-diffusion coefficient,  
$D_{\textrm{PBC}}$ is the self-diffusion coefficient obtained under periodic boundary conditions (PBC),
$\xi\approx2.837297$ for cubic simulation boxes is a constant determined by the shape of the simulation box,
$L$ is the length of the simulation box,
$\beta$ is the inverse temperature and $\eta$ is the shear viscosity.

\begin{equation}
  D_0 = D_{\textrm{PBC}} + \frac{\xi}{6\pi\beta\eta L}
  \label{eq:correction}
\end{equation}

\begin{figure}
    \centering
    \includegraphics{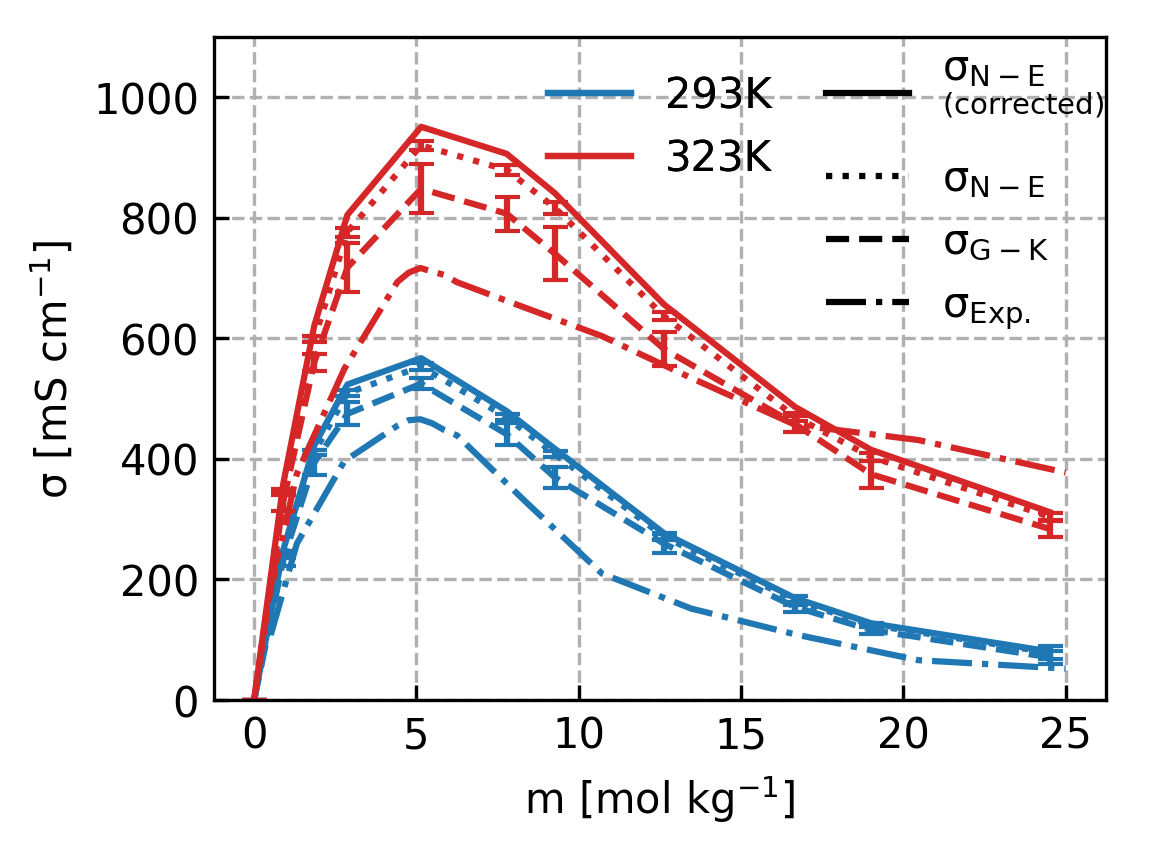}
    \caption{Ionic conductivities of NaOH solutions at 293K and 323K 
    calculated from the Nernst-Einstein formula with and without the finite-size correction,
    and that obtained from Green-Kubo formula and experimental measurements.}
    \label{fig:finite_size}
\end{figure}

We estimated the finite size effect for the Nernst-Einstein conductivity using Eq~\ref{eq:correction}
and experimental viscosity from Ref.~\citenum{visc_liquifl}. 
As is shown in Fig.~\ref{fig:finite_size}, the finite-size correction 
is rather small compared to the deviation of $\sigma_\textrm{N-E}$ from $\sigma_\textrm{G-K}$ or
$\sigma_\textrm{Exp.}$ because of the relatively large simulation box that we used and high viscosity of NaOH solutions. Therefore, we have neglected the finite-size correction and used the uncorrected 
self-diffusion coefficient and Nernst-Einstein conductivity in Fig.1 shown in the Main Text.

\section{Concentration-dependence of viscosity in NaOH solutions at 293K and 323K from experiments}
\begin{figure}[ht]
    \centering
    \includegraphics{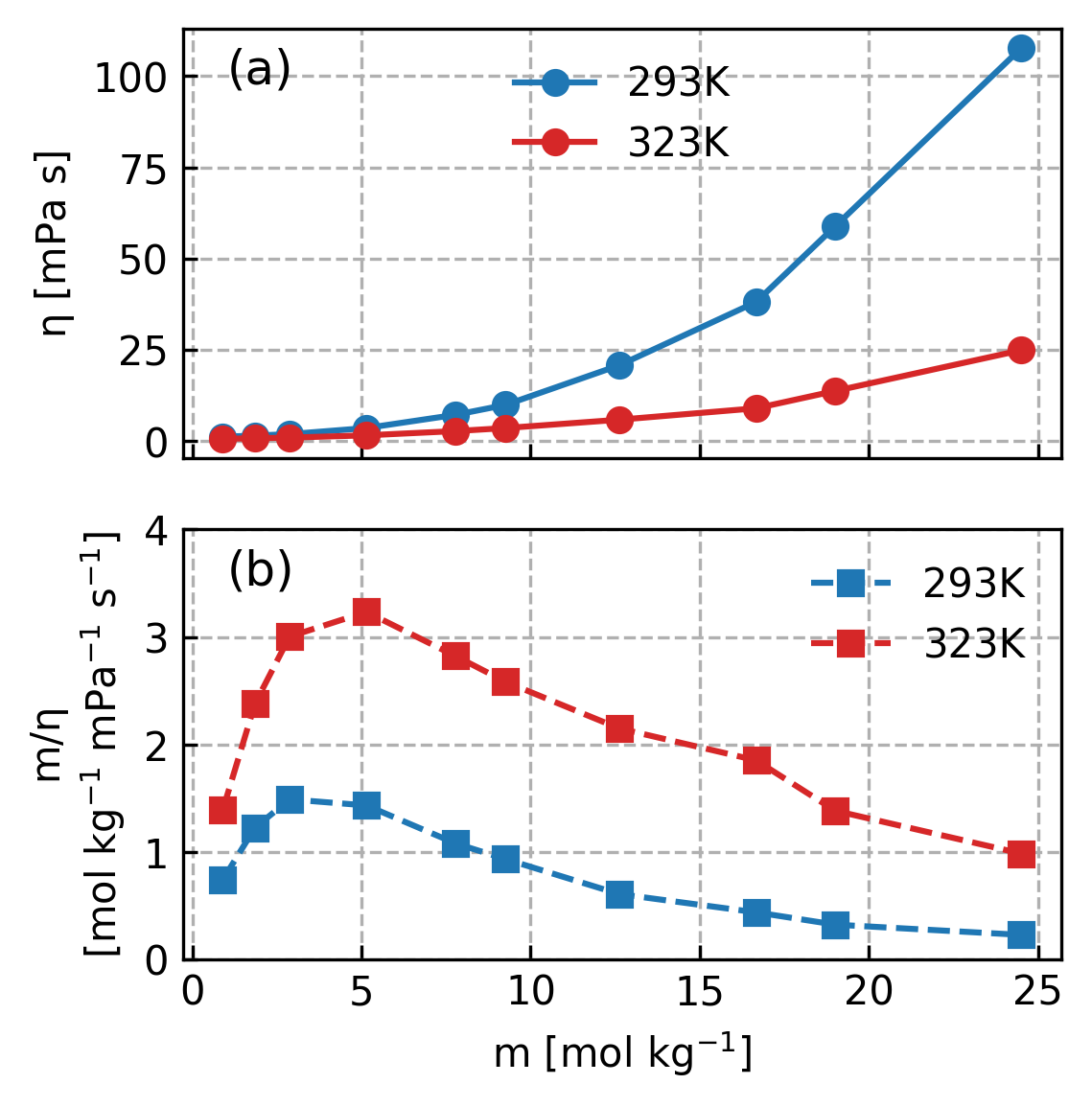}
    \caption{
    a) Concentration-dependent viscosities of NaOH solutions at 293K and 323K from
    experimental Ref.~\citenum{visc_liquifl};
    b) The corresponding concentration-weighted inverse viscosities at 293K and 323K.}
    \label{fig:exp_visc}
\end{figure}

As seen in Fig.~\ref{fig:exp_visc}a, the viscosity of NaOH solution increases rapidly with the concentration and its value is higher at 293K. 

It is interesting to note that deviations from the Nernst-Einstein relation ($\sigma_\textrm{N-E}-\sigma_\textrm{G-K}$) with a maximum around 5m are larger at 323K than those at 293K (Fig.~\ref{fig:finite_size}). This is in accord with the concentration-weighted inverse viscosity (see Fig.~\ref{fig:exp_visc}b), in the spirit of the Walden's rule~\cite{Shao2019}. 

\section{Stoichiometry of simulation boxes}

The different molalities of NaOH solutions considered in this work and the actual number of molecules
used in the simulations are given in Table~\ref{tab:stoichiometry}.

\begin{table}[]
    \centering
    \begin{tabular}{ccccccc}
    \hline
         m [mol/kg] &  $N_\mathrm{NaOH}$  &  $N_{\mathrm{H_2O}}$  &
         $L_\mathrm{293K}$ [\AA] &  $\rho_\textrm{293K}$ [g cm$^{-3}$] &
         $L_\mathrm{323K}$ [\AA] &  $\rho_\textrm{323K}$ [g cm$^{-3}$] \\
    \hline
         0.896      &  8    &  496 & 24.56 & 1.04 & 24.66 & 1.02\\
         1.852      &  16   &  480 & 24.30 & 1.07 & 24.40 & 1.06\\
         2.874      &  24   &  464 & 24.05 & 1.11 & 24.16 & 1.10\\
         5.144      &  40   &  432 & 23.59 & 1.19 & 23.70 & 1.17\\
         7.778      &  56   &  400 & 23.18 & 1.26 & 23.29 & 1.24\\
         9.259      &  64   &  384 & 22.99 & 1.30 & 23.10 & 1.28\\
         12.626     &  80   &  352 & 22.64 & 1.36 & 22.75 & 1.35\\
         16.667     &  96   &  320 & 22.34 & 1.43 & 22.45 & 1.41\\
         19.006     &  104  &  304 & 22.21 & 1.46 & 22.32 & 1.44\\
         24.509     &  120  &  272 & 21.96 & 1.52 & 22.07 & 1.50\\
    \hline
    \end{tabular}
    \caption{Compositions of simulated NaOH solutions in this work: Molality m, number of NaOH and H$_2$O molecules  $N_\mathrm{NaOH}$ and $N_\mathrm{H_2O}$, length of the cubic simulation box $L$ and
    density $\rho$ at 293K and 323K.}
    \label{tab:stoichiometry}
\end{table}

\providecommand{\latin}[1]{#1}
\makeatletter
\providecommand{\doi}
  {\begingroup\let\do\@makeother\dospecials
  \catcode`\{=1 \catcode`\}=2 \doi@aux}
\providecommand{\doi@aux}[1]{\endgroup\texttt{#1}}
\makeatother
\providecommand*\mcitethebibliography{\thebibliography}
\csname @ifundefined\endcsname{endmcitethebibliography}
  {\let\endmcitethebibliography\endthebibliography}{}